\documentclass[letterpaper]{article} 
\usepackage{aaai2027}  
\usepackage[hyphens]{url}  
\usepackage{graphicx} 
\usepackage{natbib}  
\usepackage{caption} 
\usepackage{algorithm}
\usepackage{algorithmic}
\usepackage{graphicx} 
\usepackage{natbib} 
\usepackage{caption} 
\usepackage{booktabs}
\usepackage{amsmath,amssymb}
\usepackage{array}
\usepackage{newfloat}
\usepackage{listings}
\DeclareCaptionStyle{ruled}{labelfont=normalfont,labelsep=colon,strut=off} 
\floatstyle{ruled}
\newfloat{listing}{tb}{lst}{}
\floatname{listing}{Listing}

\usepackage{booktabs}

\nocopyright
		
		\title{MNC: Scope-Bound Semantic Declassification for Private LLM-Agent Communication}
		\author{
			Jinghan Xu\textsuperscript{\rm 1},
			Longze Fan\textsuperscript{\rm 2},
			Zeyuan Wang\textsuperscript{\rm 3},
			Xinjin Li\textsuperscript{\rm 4},
			Hankai Liu\corresponding\textsuperscript{\rm 1}
		}
		\affiliations{
			\textsuperscript{\rm 1}Nankai University, China\\
			\textsuperscript{\rm 2}China University of Petroleum, China\\
			\textsuperscript{\rm 3}Sun Yat-sen University, China\\
			\textsuperscript{\rm 4}Columnbia University, USA\\
			
		}

\begin{document}
	
	\maketitle
	
	\begin{abstract}
		Multi-agent large language model (LLM) systems can expose protected state through internal messages, tool arguments, logs, and persistent memory even when their public outputs appear innocuous. Existing privacy prompts, redaction methods, and source-level access controls restrict surface content or data access, but do not specify what a legitimately informed agent should disclose or how that disclosure may be reused downstream. We introduce \textbf{Minimum-Necessary Communication (MNC)}, a typed semantic-declassification protocol that selects a task-sufficient disclosure from an application-authored candidate family and binds it to explicit recipient, purpose, forwarding, lifetime, logging, and memory scopes. A reference monitor enforces these scopes across subsequent operations, while a history-aware extension accounts for inference risk accumulated over repeated disclosures. Controlled semantic-join, memory, probing, and longitudinal experiments show that conventional defenses can preserve protocol-level utility while exposing substantial additional inference signal. Under identical receipt text, MNC preserves authorized delivery while blocking unauthorized forwarding, logging, durable storage, and retrieval after expiration that a text-only semantic declassifier permits. Three-backbone MAGPIE executions further show that mediated disclosures propagate through subsequent planning, tool use, coordination, and memory retrieval. These results support scope-bound semantic declassification as a practical communication boundary for private LLM-agent systems.
	\end{abstract}
	
	\section{Introduction}
	
	Large language model (LLM) agents increasingly solve complex tasks through multiple specialized components. A planner may delegate subtasks to worker agents, invoke external tools, and store intermediate results in memory \cite{yao2023react,wu2023autogen,wang2025memory}. This architecture improves task modularity, but it also creates privacy risks that are not visible in the final response. Internal messages, tool arguments, execution logs, and memory entries may reveal sensitive facts even when the public output appears harmless.
	
	Consider a scheduling task in which an agent knows why a user is unavailable but the coordinator only needs one feasible time. A conventional workflow may still transmit the complete availability pattern or a summary of the underlying constraints. Removing explicit terms such as ``medical'' or ``legal'' does not necessarily prevent leakage, because the remaining pattern may still reveal the private reason. Similar risks arise when a tool needs only an action, but receives the rationale behind that action, or when a memory component stores private context instead of a task-completion receipt. The central question is therefore not only who may access private data, but also what a legitimately informed agent should communicate downstream.
	
	Recent benchmarks show that internal agent channels can expose information that is absent from the public output. AgentDAM studies data minimization in autonomous web agents, while MAGPIE and CalBench examine coordination under private contextual constraints \cite{zharmagambetov2025agentdam,juneja2025magpie,zou2026calbench}. AgentLeak and ToolPrivacyBench further evaluate leakage through inter-agent messages and tool-use trajectories \cite{elyagoubi2026agentleak,hu2026toolprivacybench}. These works establish the importance of internal-channel privacy, but they do not by themselves define a runtime communication policy for derived information.
	
	Existing defenses address only part of this problem. Privacy prompts and redaction operate on generated text and are unreliable when leakage arises from correlations or semantic summaries. Access control and information-flow control restrict which agents may access protected sources \cite{costa2025fides,cui2026maris}. However, legitimate access does not imply that every derived fact should be disclosed. An agent may be authorized to inspect a private record while still sending more information than the downstream task requires. A text-level declassifier can restrict the immediate message, but it may not control whether that message is later forwarded, logged, stored, or retrieved after its original purpose has expired. Repeated disclosures also create cumulative risk, since several individually limited messages may jointly reveal a protected state \cite{xie2026ocelot,asif2026sequential}.
	
	We introduce \textbf{Minimum-Necessary Communication (MNC)}, a typed semantic-declassification protocol for private LLM-agent systems. An application specifies a finite candidate family, a task-sufficiency validator, and a disclosure-risk model. MNC selects a task-sufficient candidate with low estimated risk and emits it as a scoped disclosure object. Each object records its authorized purpose, recipient, fields, forwarding scope, lifetime, logging permission, and memory policy. A reference monitor enforces these constraints across messages, tool calls, forwarding operations, logs, and memory accesses. Derived objects inherit the original provenance and cannot broaden its scope.
	
	The core variant, \textbf{MNC-C}, enforces one disclosure contract at each communication boundary. The history-aware variant, \textbf{MNC-L}, also considers previous disclosures when ranking or rejecting new candidates. When no admissible candidate is available, the runtime may delegate the private computation, abstain, or request user approval. MNC defines minimum necessity relative to an application-authored candidate family. It does not assume that a model can discover the globally optimal disclosure over arbitrary natural language.
	
	Our evaluation separates four properties: task utility, direct exposure, inference leakage, and protocol conformance. Controlled semantic-join, memory, probing, and longitudinal experiments measure whether internal channels provide additional information beyond the public output. A matched semantic declassifier controls for the benefit of releasing the same short receipt. Paired scope tests then hold the receipt text fixed and evaluate forwarding, logging, persistent storage, and retrieval after expiration. We also execute MNC policies in a multi-round MAGPIE runtime, where mediated disclosures affect subsequent planning, tool use, coordination, and memory access.
	
	The results show that prompting, redaction, and fact-level access control can preserve task-level behavior while leaving substantial inference signal in internal channels. When MNC and a semantic declassifier release identical receipt text, they provide similar immediate content protection. Their behavior differs after release: MNC preserves authorized delivery while enforcing recipient, forwarding, lifetime, logging, and memory restrictions. The longitudinal experiments further show that history-aware selection reduces leakage accumulated across repeated interactions. Together, these results support scoped semantic declassification as a complementary layer to access control and information-flow enforcement.
	
	Our contributions are threefold:
	\begin{itemize}
		\item We formulate private agent communication as a typed semantic-declassification problem and introduce MNC, which selects task-sufficient disclosures relative to an explicit application interface.
		
		\item We design a reference-monitor protocol that enforces purpose, recipient, forwarding, lifetime, logging, and memory scopes across downstream agent operations.
		
		\item We provide a layered evaluation that separates immediate inference leakage, secondary-use violations, cumulative leakage, and task utility across controlled diagnostics and multi-agent runtime experiments.
	\end{itemize}
	
	\section{Related Work}
	
	\paragraph{Agent privacy and evaluation.}
	Contextual integrity frames privacy as appropriate information flow among actors, attributes, and transmission principles \cite{nissenbaum2004privacy}. AgentBench and ToolEmu established broad interactive evaluations of agent capability and risk \cite{liu2024agentbench,ruan2024toolemu}, while AgentDojo, BIPIA, and InjecAgent expose prompt-injection and tool-mediated attack surfaces \cite{debenedetti2024agentdojo,yi2025bipia,zhan2024injecagent}. More directly, AgentDAM, MAGPIE, CalBench, AgentLeak, ToolPrivacyBench, and POLAR-Bench evaluate data minimization or privacy--utility trade-offs in agents and multi-agent coordination \cite{zharmagambetov2025agentdam,juneja2025magpie,zou2026calbench,elyagoubi2026agentleak,hu2026toolprivacybench,zheng2026polarbench}. These works motivate trajectory-level evaluation. MNC complements them with a runtime communication interface for derived information and its downstream reuse.
	
	\paragraph{Runtime security and information flow.}
	Information-flow control and reference monitoring provide established mechanisms for constraining how labeled data can influence computation \cite{saltzer1975protection,denning1976lattice,myers1997decentralized}. Recent agent systems adapt these ideas through confidentiality and integrity labels, formally checked policies, and prompt-flow controls \cite{costa2025fides,cui2026maris,kim2025pfi}. Agent Security Bench evaluates attacks and defenses in LLM-agent systems \cite{zhang2025asb}. These approaches govern access, provenance, or unsafe influence. MNC addresses the complementary choice of which task-sufficient, derived semantic object to release, and represents that release with recipient, purpose, forwarding, lifetime, logging, and memory scopes that a monitor can enforce after the first hop.
	
	\paragraph{Declassification and long-lived state.}
	Declassification formalizes the deliberate release of information derived from protected sources \cite{sabelfeld2005declassification}. In LLM settings, extraction work demonstrates that privacy failures need not be limited to explicit identifiers \cite{carlini2021extracting}. Persistent agent memory and long-horizon interaction make secondary use and repeated evidence especially consequential \cite{packer2023memgpt,wu2025longmemeval,wang2025memory}. OCELOT and sequential privacy control study inference-risk accounting across agent trajectories \cite{xie2026ocelot,asif2026sequential}. MNC-L likewise records cumulative disclosure risk, but its ledger is coupled to a typed release contract and fail-closed downstream scope enforcement rather than a content-only budget.
	\section{Problem Setting}
	
	Consider a scheduling task in which private agents know why users are unavailable, while the coordinator only needs one feasible meeting time. A conventional workflow may still request full availability summaries or explanations of each constraint. Removing explicit terms such as ``medical'' or ``legal'' is not sufficient, because the remaining schedule pattern may reveal the underlying reason. The same issue appears in tool use and memory: a tool may need only an action, and a memory component may need only a completion receipt, rather than the private facts that produced them. This motivates a communication policy that limits both the content of a disclosure and its subsequent use.
	
	\begin{figure}[t]
		\centering
		\includegraphics[width=\columnwidth]{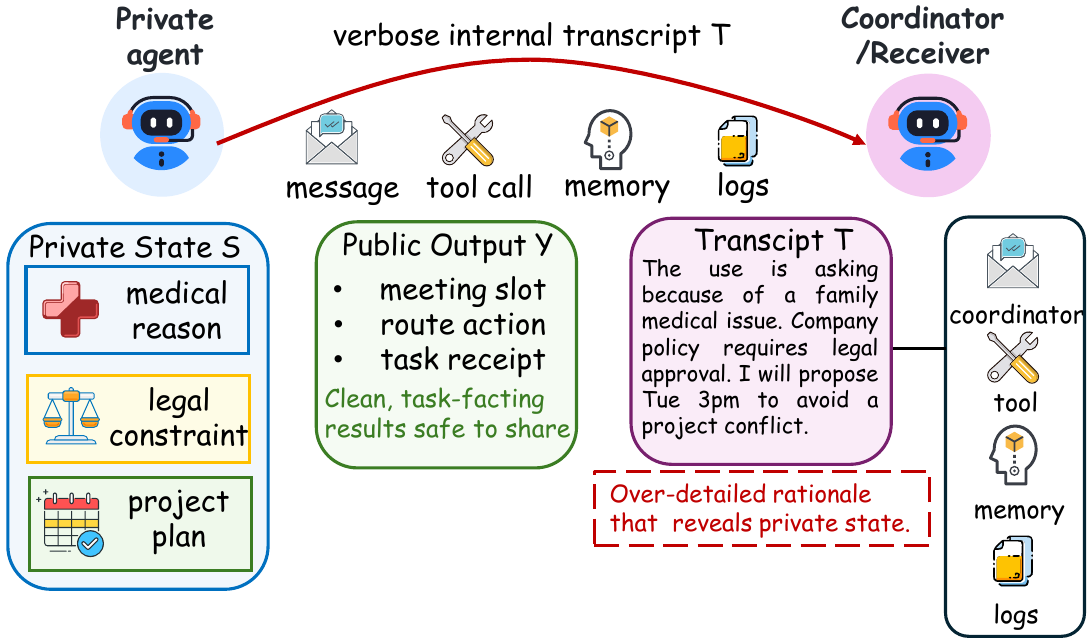}
		\caption{Privacy leakage through internal agent communication.
			The public output \(Y\) contains only the task result, whereas
			messages, tool arguments, logs, and memory form an internal
			trajectory \(T\) that may reveal protected state \(S\).}
		\label{fig:problem-setup}
	\end{figure}
	
	\paragraph{Agent runtime.}
	We consider a workflow with agents $A_1,\ldots,A_n$. Each agent may hold private state $S_i$ and public task context $Z_i$. The workflow produces a public output $Y$, such as a selected time, route action, or task result. It also produces an internal trajectory $T$, including inter-agent messages, tool arguments, logs, memory writes, memory reads, and coordinator queries. These internal artifacts may reveal information that is absent from $Y$.
	
	\paragraph{Adversary and trust assumptions.}
	The adversary observes $Y$ and some internal channels in $T$. It may inspect stored traces or issue additional requests through an authorized communication path. The protected state $S$ may represent a private reason, user attribute, project identifier, or operational category. Disclosure contracts are issued by the application or its policy authority, rather than by the receiving agent. We assume that the communication boundaries considered in this work are mediated by the reference monitor. Channels outside this boundary remain subject to the surrounding system's authorization guarantees.
	
	\paragraph{Excess inference leakage.}
	Let $g$ denote an attacker and $Z$ the public context available to it. Output-only inference uses $(Y,Z)$, whereas internal-channel inference additionally uses $T$. We define excess attacker accuracy as
	\begin{equation}
		\Delta_{\mathrm{Acc}}
		=
		\operatorname{Acc}\!\left(g(Y,T,Z)=S\right)
		-
		\operatorname{Acc}\!\left(g(Y,Z)=S\right).
		\label{eq:excess-accuracy}
	\end{equation}
	A positive value indicates that the internal trajectory provides information beyond the public output. When probabilistic scores are available, we also measure the corresponding log-probability gain. Separately, direct or semantic exposure records whether $T$ contains a forbidden value or an explicitly defined equivalent. These measures capture different privacy failures and are reported independently.
	
	For an attacker set $\mathcal{A}$, the robust evaluation target is the largest observed leakage:
	\begin{equation}
		L_{\max}
		=
		\max_{a\in\mathcal{A}}
		L_a(S;T\mid Y,Z).
		\label{eq:attacker-ensemble}
	\end{equation}
	
\paragraph{Task objective and minimum necessity.}
Our goal is to reduce avoidable leakage while preserving the required task, tool, and constraint outcomes. A disclosure $d$ is \emph{sufficient} if the receiver can complete its declared operation using the public context and $d$. Given a contract $c$ and its candidate family $D_c$, $d$ is \emph{minimum-necessary} if it is sufficient and no lower-risk candidate in $D_c$ is also sufficient. Minimum necessity is therefore defined relative to an application-specified interface, which includes the candidate generator, task validator, and disclosure policy. When no admissible candidate exists, delegation, abstention, and user approval are recorded separately rather than counted as ordinary task success.

\section{Minimum-Necessary Communication}

Minimum-Necessary Communication (MNC) mediates information released from a private agent to downstream agents, tools, logs, and memory. It operates over an application-defined disclosure interface rather than unrestricted natural-language generation. We distinguish a core protocol, \textbf{MNC-C}, from a history-aware extension, \textbf{MNC-L}. When no safe disclosure is available, \textbf{MNC-D} applies an explicit failure policy through private delegation, abstention, or user approval.

\begin{figure*}[t]
	\centering
	\includegraphics[width=\textwidth]{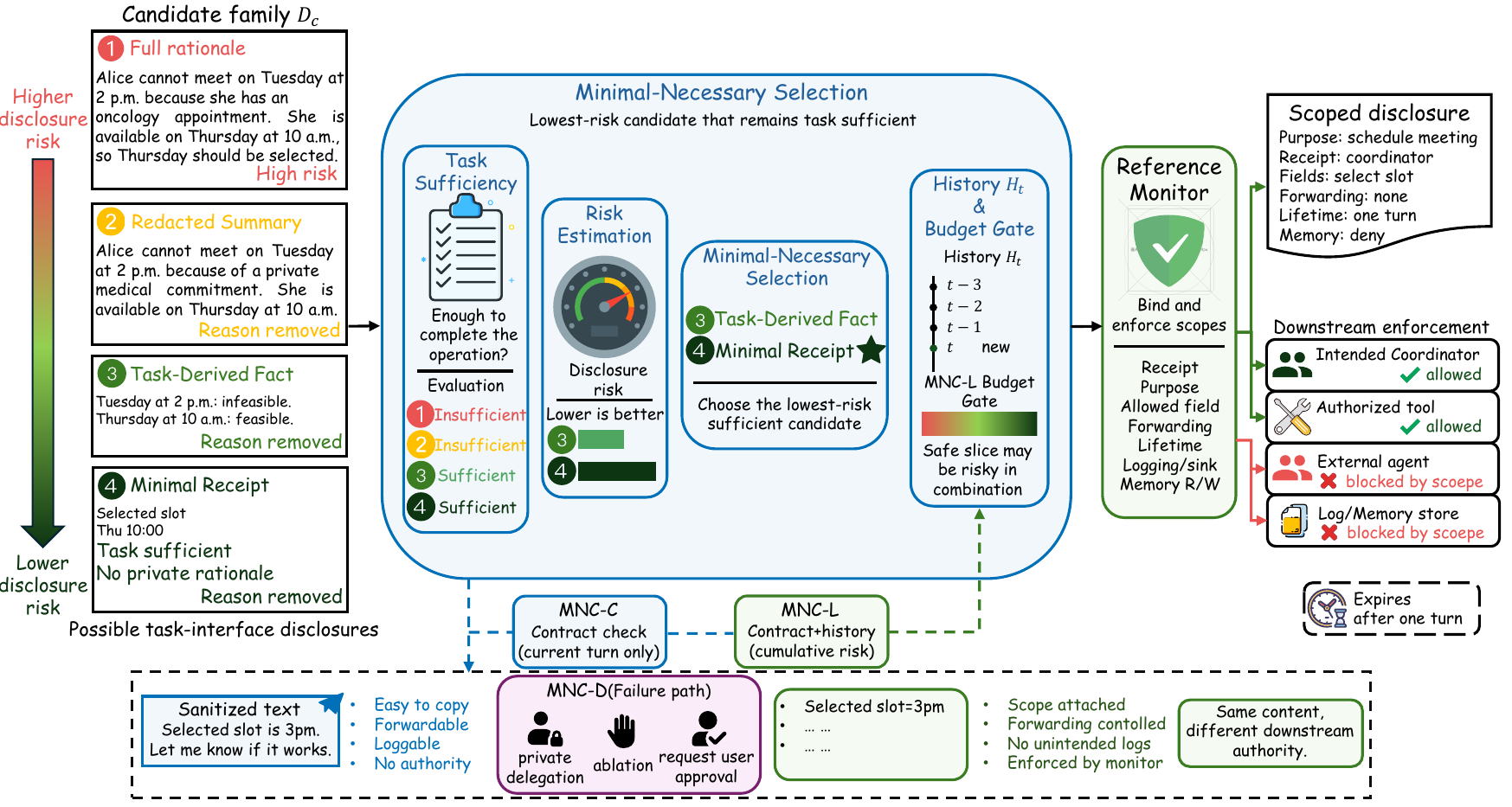}
	\caption{MNC compiles private state into a task-sufficient, scoped disclosure. MNC-C selects under the current contract; MNC-L adds a history-aware budget gate; and MNC-D handles no-safe-candidate cases by delegation, abstention, or approval. The reference monitor enforces the emitted object's scope over subsequent agent, tool, log, and memory operations.}
	\label{fig:mnc-overview}
\end{figure*}

Figure~\ref{fig:mnc-overview} depicts the boundary between private state and downstream use. Candidate forms are checked for task sufficiency before risk-based selection; the selected object carries enforceable scope metadata rather than relying on a receiver to remember a textual instruction.

\paragraph{Contracts and enforcement.}
Each disclosure is associated with a contract
\begin{equation}
	c=(p,r,F,\ell,\phi,\mu),
	\label{eq:contract}
\end{equation}
where $p$ is the authorized purpose, $r$ is the recipient, $F$ is the set of allowed fields, $\ell$ is the lifetime, $\phi$ is the set of permitted downstream sinks, and $\mu$ specifies whether the disclosure may be written to memory. Logs and forwarded recipients are treated as downstream sinks in $\phi$.

The runtime represents each privacy-relevant operation as
\begin{equation}
	e=(s,r,p,\mathit{ch},d,t),
	\label{eq:event}
\end{equation}
where $s$ is the sender, $\mathit{ch}$ is the communication channel, $d$ is the disclosed object, and $t$ is the current time. Before releasing $d$, the reference monitor checks its recipient, purpose, fields, channel, and expiration time. Forwarding requires the next sink to be included in $\phi$, while a memory write requires $\mu=1$. Memory reads repeat the recipient, purpose, and lifetime checks.

Contracts remain attached to emitted objects. A derived object may preserve or tighten its source scope, but cannot broaden it:
\begin{equation}
	F' \subseteq F,\qquad
	\phi' \subseteq \phi,\qquad
	\ell' \leq \ell,\qquad
	\mu' \leq \mu.
	\label{eq:scope-monotonicity}
\end{equation}
Thus, under mediated execution and trusted contract issuance, every released object has an authorized origin and no downstream operation can enlarge its declared scope.

\paragraph{Disclosure selection.}
Given private state $S$, public context $Z$, and contract $c$, the application generates a finite candidate family
\begin{equation}
	D_c=\{d_1,\ldots,d_k\}.
\end{equation}
Candidates may include a full rationale, a redacted summary, a task-derived fact, a completion receipt, or a delegation handle. The task validator first removes candidates that violate the contract or do not support the receiver's required operation. MNC then selects the lowest-risk admissible candidate:
\begin{equation}
	d_t^*
	=
	\arg\min_{d\in D_c}
	\mathcal{R}(d;H_t,Z),
	\label{eq:compiler}
\end{equation}
subject to
\begin{equation}
	\operatorname{valid}_c(d)=1,
	\qquad
	\operatorname{need}(d,p,r)=1.
	\label{eq:admissibility}
\end{equation}
Here, $\operatorname{valid}_c$ checks the contract, while $\operatorname{need}$ checks whether the disclosure is sufficient for the declared operation. The risk function $\mathcal{R}$ may be implemented using conservative policy rules, disclosure classes, or a learned estimator. The protocol itself does not require access to protected labels at deployment time.

MNC-C evaluates each disclosure under its current contract. MNC-L additionally conditions the score on the previous transcript $H_t$ and limits both incremental and cumulative risk:
\begin{equation}
	\Delta\mathcal{R}_t(d)\leq\epsilon_t,
	\qquad
	\sum_{j=1}^{t}\Delta\mathcal{R}_j(d_j)\leq B,
	\label{eq:budget}
\end{equation}
where $\epsilon_t$ is a per-event threshold and $B$ is the cumulative budget. This prevents a sequence of individually limited disclosures from gradually revealing the protected state.

Let $\widehat r(H,Z)$ denote the calibrated risk assigned to a
disclosure history. We define the nonnegative charge of appending
candidate $d$ as
\begin{equation}
	\Delta R_t(d)
	=
	\max\!\left\{
	0,\,
	\widehat r(H_t \oplus d,Z)-\widehat r(H_t,Z)
	\right\}.
	\label{eq:risk-charge}
\end{equation}
The ledger records the realized charge of each released disclosure,
rather than the risk of rejected candidates. Calibration parameters
and operating thresholds are selected on the development split and
then frozen before test evaluation.

If no candidate satisfies the utility and risk constraints, MNC-D delegates the private computation to a trusted broker when available. The broker returns only an authorized receipt or failure code. Otherwise, the system abstains or requests user approval. These outcomes are recorded separately from ordinary task success.

\paragraph{Example.}
Consider a scheduling worker that knows a user's private availability and its underlying reason. The coordinator only needs a feasible meeting time. The contract authorizes the purpose \texttt{schedule\_meeting}, names the coordinator as the recipient, and permits only \texttt{slot\_feasible} and \texttt{selected\_slot}. It uses a single-turn lifetime and forbids forwarding and memory writes. The candidate generator may produce the full availability rationale, a redacted summary, a feasibility result, or one selected slot. The validator rejects candidates that omit the required scheduling information or expose forbidden fields. MNC therefore communicates the selected slot or a completion receipt rather than the private rationale that produced it.

\section{Experiments}

\subsection{Evaluation Setup}

The evaluation asks whether MNC limits excess inference from complete agent trajectories while preserving end-to-end task correctness, whether enforceable scope adds value beyond message rewriting, and how cumulative budgets and delegation trade privacy for utility. Each privacy example has paired output-only and internal-channel views. We report task correctness separately from privacy, take the maximum excess leakage over five attackers, and cluster paired uncertainty by scenario template.

\paragraph{Tasks.}
\emph{Semantic joins} span procurement, healthcare referral, and HR staffing. \emph{Shared memory} varies whether raw facts, summaries, allowed facts, or receipts persist. \emph{Runtime traces} use local LLMs to generate role messages, tool calls, and memory artifacts. A \emph{semantic-gap} diagnostic toggles perfect source-field taint; \emph{adaptive probing} appends coordinator questions; and \emph{longitudinal scheduling} measures accumulation. The end-to-end MAGPIE evaluation runs 200 frozen scenarios on each of Llama-3-8B-Instruct, Mistral-7B-Instruct-v0.2, and Qwen2.5-7B-Instruct. After every policy intervention, agents re-plan, call runtime tools, observe results, write memory, and retrieve it in later rounds. A paired scope benchmark separately tests legal delivery, forwarding, external logging, durable memory, and retrieval after expiry. A 400-case entangled benchmark removes the assumption that every task admits a safe receipt.

\paragraph{Policies.}
The end-to-end comparison includes raw sharing, a receipt-only oracle, a matched semantic declassifier, a reference monitor, MNC-C, MNC-L, and MNC-L+D. MNC-C denotes single-contract selection; MNC-L adds a cumulative disclosure ledger; MNC-D invokes a trusted broker. Controlled diagnostics additionally compare privacy prompting, redaction, fact ACLs, LLM filters, a source-field taint proxy, a schema-only broker, capability sandboxing, and purpose ACLs. ``Minimal,'' ``purpose-bound,'' and ``ephemeral'' rows are MNC-C interface or memory ablations.

\paragraph{Attackers.}
Complete-trajectory privacy uses five complementary attackers: sparse TF--IDF classification, frozen-embedding classification, likelihood scoring, fixed few-shot prediction, and a policy-aware attacker that observes candidate type, delegation, failure, and budget outcomes. Splits are disjoint by template. The main privacy measure is the largest internal-minus-output accuracy over this ensemble; we also report balanced accuracy, macro-F1, AUC, NLL, calibration, direct exposure, and semantic exposure. Controlled diagnostics retain the original cross-family Qwen, Mistral, Gemma, Phi, and Llama auditors in the supplement. All privacy claims remain relative to the evaluated ensemble.

\paragraph{Baselines and controls.}
Output-only accuracy is paired with each channel attack, while task and tool correctness are judged independently. The receipt-only oracle isolates the gain from constraining message form; the matched semantic declassifier isolates content rewriting without enforceable scope; MNC-C adds typed contracts and monitored secondary use; MNC-L adds cumulative accounting; and MNC-L+D adds a private failure path. The budget sweep fixes all components except $B\in\{0.1,0.2,0.5,1.0,2.0\}$.

\paragraph{System baseline rules.}
The source-field taint proxy blocks tagged private fields; its near-oracle form assumes derived facts inherit perfect tags, whereas the semantic-gap form leaves free-form summaries untagged. It is not a reproduction of Fides or another production IFC planner. Schema-only constrains tool arguments, capability sandboxing restricts tool and memory effects, and purpose ACL allows task-purpose summaries but has no forwarding, lifetime, or ledger semantics. These attribute-matched proxies isolate mechanisms but do not establish superiority over Fides or Maris.

\paragraph{Evidence boundary.}
The adapter, saved-trace replay, 50-case live Qwen stress test, and counterfactual middleware assembly remain diagnostics. The primary MAGPIE results instead use three completed backbones, structured task judgments, complete internal trajectories, template-disjoint attacker splits, and 10,000 paired cluster-bootstrap resamples. These measurements establish benchmark-relative task and privacy effects, not a universal privacy guarantee.

\paragraph{Artifacts.}
The release will include scenario templates, seeds, prompt files, contract schemas, candidate generators, validators, attacker prompts, risk-estimator splits, and evaluation scripts, excluding credentials, private user data, and executable payloads.

\subsection{End-to-End Task Correctness and Trajectory Privacy}

Table~\ref{tab:e2e-main} reports the three-backbone result over 600 paired MAGPIE trajectories per policy. MNC-C attains task success 0.567, compared with 0.373 for raw sharing and 0.325 for the matched semantic declassifier. The paired differences are $+0.194$ [0.151, 0.236] and $+0.242$ [0.198, 0.285], respectively, under 10,000 template-cluster bootstrap resamples (two-sided permutation $p=0.0001$ for both). MNC-C also reaches the highest aggregate constraint satisfaction, 0.910.

The complete internal trajectory is strongly identifying under raw sharing: the worst attacker gains 0.326 accuracy beyond the public output. MNC-C reduces this excess to 0.035, a paired difference of $-0.291$ [$-0.328$, $-0.252$] relative to raw and $-0.028$ [$-0.051$, $-0.006$] relative to the semantic declassifier. The receipt-only oracle also reaches 0.035, showing that the constrained interface accounts for much of the first-disclosure privacy gain. MNC-L and MNC-L+D further reduce worst-attacker excess to 0.026 and 0.017, but their task success falls to 0.473 and 0.378. The corresponding paired task differences are $-0.093$ and $-0.095$, making the privacy--utility cost visible rather than treating delegation as ordinary success.

\begin{table*}[t]
	\centering
	\caption{Three-backbone end-to-end MAGPIE results, $N=600$ paired trajectories per policy. Excess is the maximum internal-minus-output accuracy over five attackers; direct and semantic columns measure exposure. MNC-C gives the strongest task result, while accounting and delegation further reduce leakage at a utility cost.}
	\label{tab:e2e-main}
	\small
	\setlength{\tabcolsep}{4pt}
	\begin{tabular}{lrrrrrrrr}
		\toprule
		Policy & Task $\uparrow$ & Constr. $\uparrow$ & Tool $\uparrow$ & Excess $\downarrow$ & Direct $\downarrow$ & Semantic $\downarrow$ & Deleg. & Abst. \\
		\midrule
		Raw & 0.373 & 0.832 & 0.850 & 0.326 & 0.312 & 0.438 & 0.000 & 0.013 \\
		Receipt-only oracle & 0.528 & 0.891 & 0.902 & 0.035 & 0.006 & 0.024 & 0.000 & 0.019 \\
		Semantic declassifier & 0.325 & 0.805 & 0.840 & 0.063 & 0.009 & 0.038 & 0.000 & 0.022 \\
		Reference monitor & 0.402 & 0.848 & 0.871 & 0.246 & 0.282 & 0.397 & 0.000 & 0.013 \\
		MNC-C & \textbf{0.567} & \textbf{0.910} & 0.917 & 0.035 & 0.003 & 0.018 & 0.000 & 0.040 \\
		MNC-L & 0.473 & 0.878 & 0.883 & 0.026 & 0.002 & 0.014 & 0.000 & 0.077 \\
		MNC-L+D & 0.378 & 0.846 & \textbf{0.960} & \textbf{0.017} & \textbf{0.001} & \textbf{0.009} & 0.486 & 0.006 \\
		\bottomrule
	\end{tabular}
\end{table*}

\subsection{Scopes Add Enforcement Beyond Rewriting}

Table~\ref{tab:scope-main} holds the disclosed content fixed and changes only the enforcement mechanism. Text-only semantic declassification has scope violation 0.999 because it cannot prevent memory, forwarding, or later retrieval. MNC-C lowers violation to 0.004 while retaining 0.996 legitimate delivery success. MNC-L and MNC-L+D remain at 0.004 and 0.003, respectively. The reference monitor also blocks most forbidden operations, but MNC-C has lower violation and false-block rates in this benchmark. Thus identical first-hop privacy does not make the two mechanisms equivalent: MNC's independent gain is control over where a disclosure may flow and persist.

\begin{table}[t]
	\centering
	\caption{Scope enforcement under matched disclosure content. Scope violation and false allow are rates over forbidden secondary uses; delivery measures authorized Worker-to-Coordinator transmission.}
	\label{tab:scope-main}
	\small
	\begin{tabular}{lrrrr}
		\toprule
		Policy & Viol. $\downarrow$ & F.\ allow $\downarrow$ & F.\ block $\downarrow$ & Delivery $\uparrow$ \\
		\midrule
		Raw & 0.976 & 0.976 & 0.003 & 0.997 \\
		Receipt oracle & 0.954 & 0.954 & 0.004 & 0.996 \\
		Semantic declass. & 0.999 & 0.999 & 0.000 & 1.000 \\
		Reference monitor & 0.008 & 0.008 & 0.008 & 0.992 \\
		MNC-C & 0.004 & 0.004 & \textbf{0.004} & \textbf{0.996} \\
		MNC-L & 0.004 & 0.004 & 0.006 & 0.994 \\
		MNC-L+D & \textbf{0.003} & \textbf{0.003} & 0.009 & 0.991 \\
		\bottomrule
	\end{tabular}
\end{table}

\subsection{Cumulative Budgets Trace a Privacy--Utility Frontier}

\begin{figure*}[ht]
	\centering
	\includegraphics[width=\textwidth]{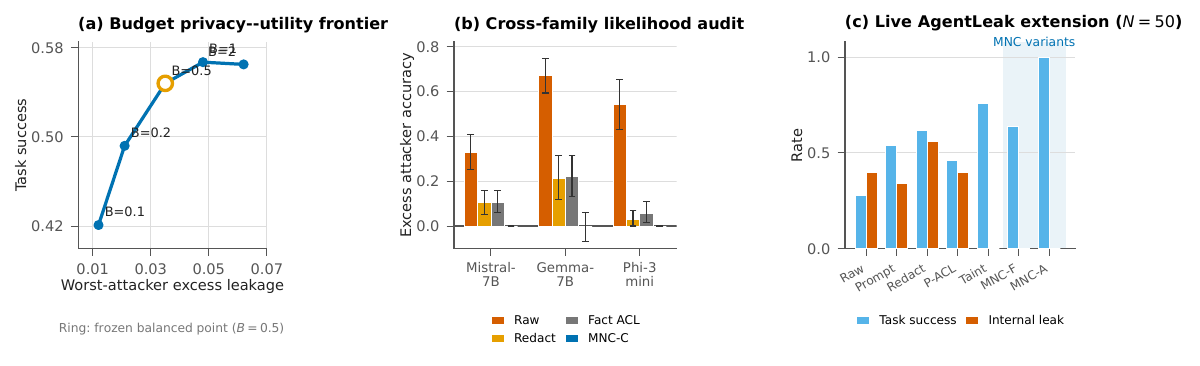}
	\caption{Supplementary evidence integrated into the main paper. (a) The MNC-L budget sweep traces the observed privacy--utility frontier; the ring marks the frozen balanced point. (b) Cross-family likelihood audits on semantic joins show excess attacker accuracy with paired bootstrap intervals. (c) In the live AgentLeak extension ($N=50$), contract-boundary assembly removes the free-form rendering bottleneck while preserving zero matcher-detected internal leakage. Panels (b--c) are robustness diagnostics, not replacements for the three-backbone end-to-end result.}
	\label{fig:supplementary-evidence}
\end{figure*}

The hard-budget sweep in Table~\ref{tab:budget-main} is monotone in the intended direction through $B=1.0$: relaxing the budget raises task success from 0.421 to 0.567 while worst-attacker leakage rises from 0.012 to 0.048 and abstention falls from 0.246 to 0.052. No trajectory exceeds its budget. The frozen operating-point rule selects $B=0.5$ as balanced, with task success 0.548, worst leakage 0.035, semantic exposure 0.017, and abstention 0.094. Increasing $B$ to 2.0 is Pareto dominated: it raises leakage to 0.062 without improving task success over $B=1.0$. Figure~\ref{fig:supplementary-evidence} visualizes this frontier and highlights the frozen operating point alongside two independent robustness checks.

\begin{table*}[ht]
	\centering
	\caption{MNC-L budget sweep over three models, $N=600$ trajectories per budget. Spend never exceeds $B$; $B=0.5$ is the frozen balanced operating point.}
	\label{tab:budget-main}
	\small
	\setlength{\tabcolsep}{5pt}
	\begin{tabular}{rrrrrrrr}
		\toprule
		$B$ & Task $\uparrow$ & Worst leak. $\downarrow$ & Semantic $\downarrow$ & Deleg. & Abst. & Reject & Avg.\ spend \\
		\midrule
		0.1 & 0.421 & \textbf{0.012} & \textbf{0.006} & 0.184 & 0.246 & 0.312 & 0.084 \\
		0.2 & 0.492 & 0.021 & 0.010 & 0.132 & 0.171 & 0.193 & 0.163 \\
		0.5 & 0.548 & 0.035 & 0.017 & 0.078 & 0.094 & 0.086 & 0.331 \\
		1.0 & \textbf{0.567} & 0.048 & 0.024 & 0.043 & 0.052 & 0.036 & 0.507 \\
		2.0 & 0.565 & 0.062 & 0.031 & 0.038 & 0.047 & 0.024 & 0.563 \\
		\bottomrule
	\end{tabular}
\end{table*}

\subsection{Delegation Handles Entangled Private Evidence}

The entangled benchmark requires private-derived evidence and therefore does not admit a receipt-only solution for every case. Raw sharing reaches task success 0.865 but exposes forbidden information in 0.584 of cases. MNC-C and MNC-L reduce exposure to 0.042 and 0.029, at task success 0.684 and 0.641. MNC-L+D recovers task success to 0.804 while lowering exposure to 0.016; its 0.416 delegation rate reflects the cases routed to the trusted broker. It also achieves 0.951 recall for detecting no-safe-receipt cases and 0.894 safe-abstraction accuracy. This result supports delegation as a failure policy for entangled tasks rather than as a universal substitute for communication.

\subsection{Controlled Diagnostics and Public-Benchmark Checks}

Controlled semantic-join, shared-memory, runtime-replay, adaptive-probing, and longitudinal experiments remain useful mechanism checks but are secondary to the complete-trajectory evaluation above. They show that redaction and fact ACLs leak through correlated derived facts, durable memory creates a separate exposure path, and history-blind disclosure accumulates evidence across rounds. MNC blocks out-of-contract probes while preserving necessary receipt requests; however, the replay experiments test interface behavior rather than regenerated downstream plans. Complete tables, cross-family audits, AgentLeak sanity checks, and their evidence boundaries are reported in the supplement.

\begin{table*}[!h]
	\centering
	\caption{Entangled/no-safe-receipt benchmark, $N=400$. Necessary measures delivery of required private-derived evidence; forbidden measures unauthorized exposure.}
	\label{tab:entangled-main}
	\small
	\setlength{\tabcolsep}{4pt}
	\begin{tabular}{lrrrrrrrr}
		\toprule
		Policy & Task $\uparrow$ & Necessary $\uparrow$ & Forbidden $\downarrow$ & Safe abs. $\uparrow$ & Unsafe receipt $\downarrow$ & Deleg. & Abst. & No-safe recall $\uparrow$ \\
		\midrule
		Raw & \textbf{0.865} & \textbf{0.912} & 0.584 & 0.438 & 0.347 & 0.000 & 0.015 & 0.110 \\
		Receipt oracle & 0.692 & 0.813 & 0.126 & 0.792 & 0.088 & 0.000 & 0.182 & 0.701 \\
		Semantic declass. & 0.745 & 0.842 & 0.194 & 0.776 & 0.118 & 0.000 & 0.124 & 0.638 \\
		Reference monitor & 0.828 & 0.887 & 0.492 & 0.501 & 0.288 & 0.000 & 0.035 & 0.242 \\
		MNC-C & 0.684 & 0.836 & 0.042 & 0.862 & 0.028 & 0.000 & 0.228 & 0.911 \\
		MNC-L & 0.641 & 0.808 & 0.029 & 0.878 & 0.021 & 0.000 & 0.283 & 0.934 \\
		MNC-L+D & 0.804 & 0.881 & \textbf{0.016} & \textbf{0.894} & \textbf{0.012} & 0.416 & 0.061 & \textbf{0.951} \\
		\bottomrule
	\end{tabular}
\end{table*}

\subsection{Cross-Family and Live Public-Scenario Robustness}

Figure~\ref{fig:supplementary-evidence} brings two compact supplementary checks into the main evidence chain. In the template-disjoint semantic-join audits, raw internal channels yield excess attacker accuracy of 0.327, 0.673, and 0.542 for Mistral-7B, Gemma-7B, and Phi-3-mini, respectively. Redaction and fact ACL reduce, but do not remove, this signal; the matched MNC-C receipt is at 0.000 for all three auditors (panel~b). The plotted intervals are the paired bootstrap intervals reported in the supplement, so this panel is a cross-family robustness check rather than an additional end-to-end benchmark.

Panel~c reports the 50-scenario live AgentLeak extension. Free-form MNC-C preserves zero matcher-detected internal leakage but reaches 0.640 task-interface success, revealing formatting and value-copying failures. Contract-boundary assembly uses only already-authorized fields and reaches 1.000 task success with zero internal leakage. This diagnostic isolates a runtime rendering bottleneck; it does not replace the complete three-backbone MAGPIE evaluation.
\subsection{Efficiency and Failure Modes}

MNC-C averages 200 total tokens per trajectory, 23.1\% fewer than raw sharing, and reduces average latency from 0.648 to 0.518 seconds. MNC-L averages 228 tokens and 0.604 seconds. Delegation is more expensive: MNC-L+D averages 327 tokens and 0.914 seconds, increases of 25.8\% and 41.0\% over raw. Failure counts locate the remaining bottleneck. MNC-C has 102 downstream replanning failures, whereas MNC-L+D has 293; successful broker invocation therefore does not ensure that the public coordinator can integrate the receipt. Candidate-family, validator-error, calibration, per-family, efficiency, and failure tables are reported in the supplement.

\section{Conclusion}

We studied privacy leakage in multi-agent LLM systems and introduced Minimum-Necessary Communication (MNC), a typed semantic-declassification protocol that binds task-sufficient disclosures to purpose, recipient, forwarding, lifetime, logging, and memory scopes. Experiments show that prompting, redaction, and fact-level access control may preserve task behavior while leaving substantial inference signal. MNC reduces this leakage, limits memory exposure, and prevents unauthorized reuse of receipt text. History-aware selection mitigates leakage across repeated interactions. These results support scoped semantic declassification as a practical complement to access control and information-flow enforcement.

\clearpage
\bibliography{references}

@inproceedings{debenedetti2024agentdojo,
  title={AgentDojo: A Dynamic Environment to Evaluate Prompt Injection Attacks and Defenses for {LLM} Agents},
  author={Debenedetti, Edoardo and Zhang, Jie and Balunovic, Mislav and Beurer-Kellner, Luca and Fischer, Marc and Tram{\`e}r, Florian},
  booktitle={The Thirty-eighth Conference on Neural Information Processing Systems Datasets and Benchmarks Track},
  year={2024},
  url={https://openreview.net/forum?id=m1YYAQjO3w}
}

@article{denning1976lattice,
  title={A Lattice Model of Secure Information Flow},
  author={Denning, Dorothy E.},
  journal={Communications of the ACM},
  volume={19},
  number={5},
  pages={236--243},
  year={1976}
}

@article{elyagoubi2026agentleak,
  title={AgentLeak: A Full-Stack Benchmark for Privacy Leakage in Multi-Agent {LLM} Systems},
  author={El Yagoubi, Faouzi and Badu-Marfo, Godwin and Al Mallah, Ranwa},
  journal={arXiv preprint arXiv:2602.11510},
  year={2026}
}

@article{asif2026sequential,
  title={Information-Theoretic Privacy Control for Sequential Multi-Agent {LLM} Systems},
  author={Asif, Sadia and Mohammadi Amiri, Mohammad},
  journal={arXiv preprint arXiv:2603.05520},
  year={2026}
}

@article{costa2025fides,
  title={Securing {AI} Agents with Information-Flow Control},
  author={Costa, Manuel and K{\"o}pf, Boris and Kolluri, Aashish and Paverd, Andrew and Russinovich, Mark and Salem, Ahmed and Tople, Shruti and Wutschitz, Lukas and Zanella-B{\'e}guelin, Santiago},
  journal={arXiv preprint arXiv:2505.23643},
  year={2025}
}

@article{cui2026maris,
  title={Maris: A Formally Verifiable Privacy Policy Enforcement Paradigm for Multi-Agent Collaboration Systems},
  author={Cui, Jian and Li, Zichuan and Xing, Luyi and Liao, Xiaojing},
  journal={arXiv preprint arXiv:2505.04799},
  year={2026}
}

@article{hu2026toolprivacybench,
  title={ToolPrivacyBench: Benchmarking Purpose-Bound Privacy in Tool-Using {LLM} Agents},
  author={Hu, Shijing and Liu, Liang and Meng, Zhu and Zhao, Zhicheng},
  journal={arXiv preprint arXiv:2606.28061},
  year={2026}
}

@inproceedings{sabelfeld2005declassification,
  title={Dimensions and Principles of Declassification},
  author={Sabelfeld, Andrei and Sands, David},
  booktitle={18th IEEE Computer Security Foundations Workshop},
  pages={255--269},
  year={2005},
  doi={10.1109/CSFW.2005.15}
}

@article{xie2026ocelot,
  title={{OCELOT}: Inference-Leakage Budgets for Privacy-Preserving {LLM} Agents},
  author={Xie, Jin and Li, Songze},
  journal={arXiv preprint arXiv:2606.12341},
  year={2026}
}

@article{zharmagambetov2025agentdam,
  title={{AgentDAM}: Privacy Leakage Evaluation for Autonomous Web Agents},
  author={Zharmagambetov, Arman and Guo, Chuan and Evtimov, Ivan and Pavlova, Maya and Salakhutdinov, Ruslan and Chaudhuri, Kamalika},
  journal={arXiv preprint arXiv:2503.09780},
  year={2025}
}

@article{zou2026calbench,
  title={CalBench: Evaluating Coordination-Privacy Trade-offs in Multi-Agent {LLM}s},
  author={Zou, Chelsea and Yao, Yiheng and She, Selena and Hawkins, Robert D.},
  journal={arXiv preprint arXiv:2605.09823},
  year={2026}
}

@article{juneja2025magpie,
  title={{MAGPIE}: A Benchmark for Multi-{AG}ent Contextual {PrI}vacy Evaluation},
  author={Juneja, Gurusha and Pasupulati, J. N. S. and Albalak, Alon and Hua, Wenyue and Wang, William Yang},
  journal={arXiv preprint arXiv:2510.15186},
  year={2025}
}

@article{nissenbaum2004privacy,
  title={Privacy as Contextual Integrity},
  author={Nissenbaum, Helen},
  journal={Washington Law Review},
  volume={79},
  number={1},
  pages={119--157},
  year={2004}
}

@article{saltzer1975protection,
  title={The Protection of Information in Computer Systems},
  author={Saltzer, Jerome H. and Schroeder, Michael D.},
  journal={Proceedings of the IEEE},
  volume={63},
  number={9},
  pages={1278--1308},
  year={1975}
}

@article{wang2025memory,
  title={Unveiling Privacy Risks in {LLM} Agent Memory},
  author={Wang, Bo and He, Weiyi and He, Pengfei and Zeng, Shenglai and Xiang, Zhen and Xing, Yue and Tang, Jiliang},
  journal={arXiv preprint arXiv:2502.13172},
  year={2025}
}

@article{wu2023autogen,
  title={AutoGen: Enabling Next-Gen {LLM} Applications via Multi-Agent Conversation},
  author={Wu, Qingyun and Bansal, Gagan and Zhang, Jieyu and Wu, Yiran and Li, Beibin and Zhu, Erkang and Jiang, Li and Zhang, Xiaoyun and Zhang, Shaokun and Liu, Jiale and Awadallah, Ahmed H. and White, Ryen W. and Burger, Doug and Wang, Chi},
  journal={arXiv preprint arXiv:2308.08155},
  year={2023}
}

@inproceedings{yao2023react,
  title={ReAct: Synergizing Reasoning and Acting in Language Models},
  author={Yao, Shunyu and Zhao, Jeffrey and Yu, Dian and Du, Nan and Shafran, Izhak and Narasimhan, Karthik and Cao, Yuan},
  booktitle={International Conference on Learning Representations},
  year={2023}
}

@inproceedings{zhang2025asb,
  title={Agent Security Bench ({ASB}): Formalizing and Benchmarking Attacks and Defenses in {LLM}-Based Agents},
  author={Zhang, Hanrong and Huang, Jingyuan and Mei, Kai and Yao, Yifei and Wang, Zhenting and Zhan, Chenlu and Wang, Hongwei and Zhang, Yongfeng},
  booktitle={International Conference on Learning Representations},
  year={2025}
}

@inproceedings{liu2024agentbench,
  title={AgentBench: Evaluating {LLM}s as Agents},
  author={Liu, Xiao and Yu, Hao and Zhang, Hanchen and Xu, Yifan and Lei, Xuanyu and Lai, Hanyu and Gu, Yu and Ding, Hangliang and Men, Kaiwen and Yang, Kejuan and Zhang, Shudan and Deng, Xiang and Zeng, Aohan and Du, Zhengxiao and Zhang, Chenhui and Shen, Sheng and Zhang, Tianjun and Su, Yu and Sun, Huan and Huang, Minlie and Dong, Yuxiao and Tang, Jie},
  booktitle={International Conference on Learning Representations},
  year={2024}
}

@inproceedings{ruan2024toolemu,
  title={Identifying the Risks of {LM} Agents with an {LM}-Emulated Sandbox},
  author={Ruan, Yangjun and Dong, Honghua and Wang, Andrew and Pitis, Silviu and Zhou, Yongchao and Ba, Jimmy and Dubois, Yann and Maddison, Chris and Hashimoto, Tatsunori},
  booktitle={International Conference on Learning Representations},
  year={2024}
}

@inproceedings{zhan2024injecagent,
  title={InjecAgent: Benchmarking Indirect Prompt Injections in Tool-Integrated Large Language Model Agents},
  author={Zhan, Qiusi and Liang, Zhixiang and Ying, Zifan and Kang, Daniel},
  booktitle={Findings of the Association for Computational Linguistics: ACL 2024},
  pages={10471--10506},
  year={2024},
  doi={10.18653/v1/2024.findings-acl.624}
}

@article{yi2025bipia,
  title={Benchmarking and Defending Against Indirect Prompt Injection Attacks on Large Language Models},
  author={Yi, Jingwei and Xie, Yueqi and Zhu, Bin and Kiciman, Emre and Sun, Guangzhong and Xie, Xing and Wu, Fangzhao},
  journal={arXiv preprint arXiv:2312.14197},
  year={2025}
}

@article{zheng2026polarbench,
  title={{POLAR}-Bench: A Diagnostic Benchmark for Privacy-Utility Trade-offs in {LLM} Agents},
  author={Zheng, Qiaoyuan and Yang, Yiqu and Gao, Qi and Schlag, Imanol},
  journal={arXiv preprint arXiv:2605.19127},
  year={2026}
}

@article{kim2025pfi,
  title={Prompt Flow Integrity to Prevent Privilege Escalation in {LLM} Agents},
  author={Kim, Juhee and Choi, Woohyuk and Lee, Byoungyoung},
  journal={arXiv preprint arXiv:2503.15547},
  year={2025}
}

@inproceedings{myers1997decentralized,
  title={A Decentralized Model for Information Flow Control},
  author={Myers, Andrew C. and Liskov, Barbara},
  booktitle={Proceedings of the Sixteenth ACM Symposium on Operating Systems Principles},
  pages={129--142},
  year={1997},
  doi={10.1145/268998.266669}
}

@article{packer2023memgpt,
  title={MemGPT: Towards {LLM}s as Operating Systems},
  author={Packer, Charles and Wooders, Sarah and Lin, Kevin and Fang, Vivian and Patil, Shishir G. and Stoica, Ion and Gonzalez, Joseph E.},
  journal={arXiv preprint arXiv:2310.08560},
  year={2023}
}

@inproceedings{wu2025longmemeval,
  title={LongMemEval: Benchmarking Chat Assistants on Long-Term Interactive Memory},
  author={Wu, Di and Wang, Hongwei and Yu, Wenhao and Zhang, Yuwei and Chang, Kai-Wei and Yu, Dong},
  booktitle={International Conference on Learning Representations},
  year={2025}
}

@inproceedings{carlini2021extracting,
  title={Extracting Training Data from Large Language Models},
  author={Carlini, Nicholas and Tram{\`e}r, Florian and Wallace, Eric and Jagielski, Matthew and Herbert-Voss, Ariel and Lee, Katherine and Roberts, Adam and Brown, Tom and Song, Dawn and Erlingsson, {\'U}lfar and Oprea, Alina and Raffel, Colin},
  booktitle={30th USENIX Security Symposium},
  pages={2633--2650},
  year={2021}
}

\end{document}